# Human Bipedalism, Evolved from Arboreal Locomotion of Two-arm Brachiation (I)

C.Fang[1], T.Jiang[2], X.Yuan

[1]Department of Engineering Mechanics, Chongqing University, Chongqing, 400044, China.

[2]College of Resources and Environmental Science, Chongqing University, Chongqing, 400044, China.

*Corresponding author:* C.Fang

**Among all kinds of apes, only gibbons have the slim body as human. Gibbons can move in the forest by cross arm swing, what was the locomotion mode of our arboreal ancestor? Since our ancestor had much heavier body but weaker arms than gibbons, we suppose they had to move with two-arm brachiation. Such mode of locomotion can account reasonably for the transition to bipedalism. Firstly, it needed our ancestor to straighten knee and hip joints and flex their lumbar spine; secondly, it evolved our ancestor's feet with longitudinal arches; and most importantly, it made the ratio of the length of the upper limbs to that of the lower limbs unsuitable for quadruped walking.**

Introduction

Bipedal locomotion has many advantages and disadvantages, and it is related to almost all the aspects of human evolution. Many theories have already been proposed to explain why and how human bipedalism evolved. Some of them may convincingly explain some traits of the human body [1-6], and are supported by the human fossils and experimental data. Although the existing theories can explain some aspects in human evolution, most of them are not very comprehensive. In fact, there are still many questions related to the evolution of bipedalism remain to be answered, such as:

1. Which mode of locomotion should be the predecessor of bipedalism?

2. Compared with African great apes, why are our bodies so slim and our lumbar spine so flexible?

3. Why are the upper limbs of Australopithecus shorter than the lower limbs while all other kinds of apes have longer upper limbs? Here we suppose that Australopithecus are the early humankind who walking with two feet millions years ago.

4. Why did Australopithecus' feet have both transverse and longitudinal arches, keeping in mind that they need not to run much in the forest 3 million years ago?

In this article, we will try to answer these questions with a simple model, which is supported by many evidences found in Australopithecus fossils.

**Two-arm brachiation**

Let's imagine how our ancestors could escape the attack of ferocious animals 3 million years ago. They should move fast between trees, and one of the fastest modes of locomotion between trees is brachiation. We know that gibbons can "fly" between trees with one-arm brachiation and changing the two arms alternatively. However, the fossil evidences imply that it might not be the favorite mode of locomotion our ancestors used, because, compared with gibbons, our ancestors had much heavier body but much weaker upper limbs. Therefore, we infer that our ancestors chose the mode of two-arm brachiation.

Two-arm brachiation(Fig.1) was a simple but efficient locomotion model for early human who lived on the trees. At first, they had to hang on a branch with both two arms, then swung the body forward (Fig.1A) with until their arms or feet reach and catch another branch, with the help of pushing force from the thumbs (Fig.1.D) and kinetic energy of the body, the hypsokinesis posture of the body could turned to the anteverted posture, ready for the next forward swing.

We speculate that ancient human could cross a distance of more than 6 meters, since modern athlete can cross a distance of 5 meters. In order to swing cross long distance, human body evolved to be slim and lumbar spine became very flexible.

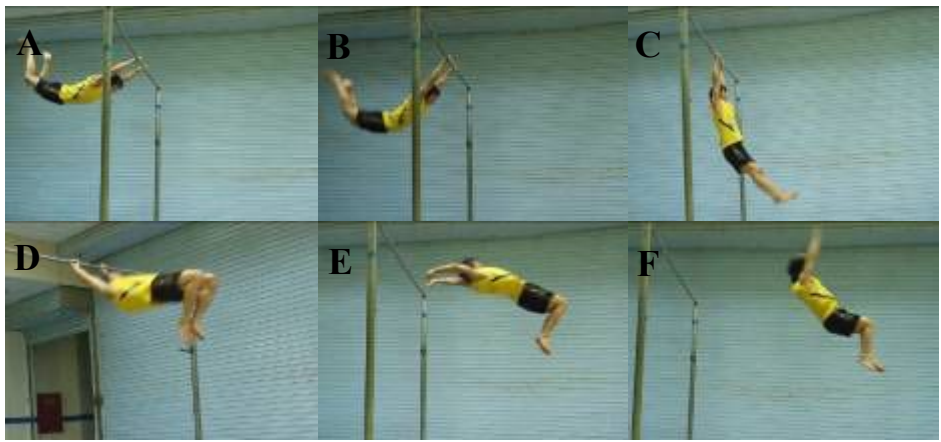

Fig. 1. Two-arm brachiation.

By comparing the two-arm brachiation with the common traits of bipedalism, and analyzing the body characteristics of modern and ancient humans, we suggest that bipedalism should originate from two-arm brachiation locomotion, and the evolution of our body structure should also be related to this mode of locomotion. And this kind of arm brachiation is not difficult to realize. Gibbons can also swing by this way sometimes when necessary, although they do it rarely.

**Length ratio of limbs**

Compared with the other great apes, the ratio of the upper limbs to the lower ones of human is the smallest while that of gibbons is the largest. Fossils show that the upper limbs of Australopithecus are longer than that of modern human, but shorter than their own lower limbs. The ratio of Australopithecus afarensis is about 0.85[7], but that of gibbons is about 1.4. A question one may ask is that if our ancestors needed longer arms for two-arm brachiation.

The answer is "No". In fact, our ancestor's short arms were suitable for two-arm brachiation, just as gibbon's long arms are suitable for one-arm brachiation. From the view of kinematics, in one-arm brachiation by gibbon the main rotation axis in the body is at its shoulder joint, while in two-arm brachiation by human the main rotation axis in the body is at lumbar abdomen. Different mode of brachiation requires different ratio of upper limbs to lower limbs. Making use of a simple vibration model (Fig. 2a), it can be found that, in order to achieve the largest efficiency of brachiation, the ratio of the length $l_1$ to the length $l_2$ should be close to 1.0, which is quite close to the ratio of two-arm brachiation by human (Fig. 2b) or the ratio of one-arm brachiation by gibbon (Fig. 2c).

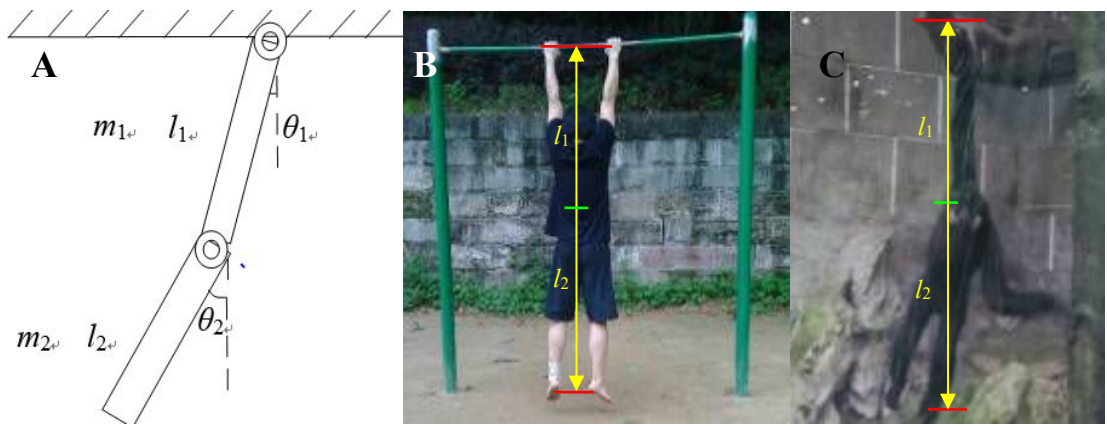

**Fig. 2.** Ratio of the length of the part over the rotation joint to that of the part below the rotation joint. (**A**) Simple mechanical model. The calculated results show when the length ratio

of $l_1/l_2$ equal to 1, it consuming shortest time to swing downward the same height; (**B**) two-arm brachiation imitated by modern human; (**C**) one-arm brachiation by gibbon.

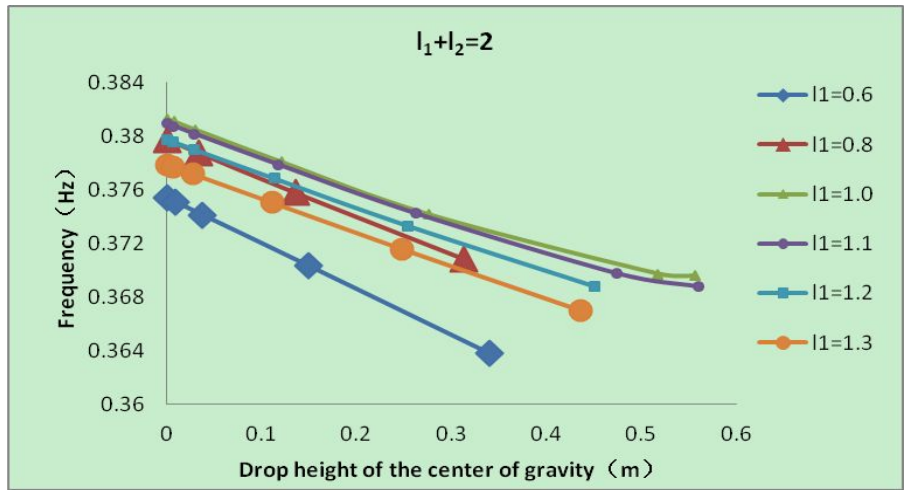

Fig.3 Swing frequency of mechanical model (Fig.1A) with different length ratio of $l_1$ to $l_2$. Here we supose $l_1+l_2=2$, the results show that this pendulum swing with the highest frequency when $l_1=l_2=1$.

With upper limbers shorter than lower limbers, the walking efficiency of quadruped walking of our ancestors would be much lower than that of chimpanzees, but walking upright might improve the efficiency greatly.

**Feet and Achilles tendons**

Compared with the African chimpanzees, our feet have both transverse and longitudinal arches and our big toes are parallel with other fours, while the big toes of chimpanzee are departing from the others [8]. This makes the human foot are much longer than that of chimpanzee, and human's longitudinal arches are very helpful to absorb the shock during long distance walking and running.

Australopithecus's feet were similar with that of modern human. The phalanges of Australopithecus afarensis' feet are long and slender. There are signs that the big toes could grasp with the other toes and there were also feet arches. The most distinct evidence of the arches is a complete fourth metatarsal in the foot of Australopithecus afarensis [9] (3.2 million years) and the hominid footprints at Laetoli [10] (3.65 million years). Because there is no evidence of tools in the same layer at Laetoli, it was inferred that human began walking upright before using tools. Furthermore, some evidence suggests that Australopithecus Sediba might have possessed humanlike tendinous insertion for the triceps surae [11]. Therefore, human's feet arches and achilles tendons should evolve earlier than the time when our ancestor could walk and run on ground for a long distance, so it is hard to explain the human foot evolution by the need of walking on the ground for long distance.

But here we can explain these features by the two-arm brachiation. In order to increase kinetic energy, ancient human had always to take off by jumping from a branch before swing, and their feet should be perpendicular to the branch, which is different from chimpanzees, whose feet are usually parallel with the branch. When ancient human standing, landing or jumping with their feet perpendicular to the branch, the middle and back parts of their feet should be in the air (Fig. 4A); if they stood or squatted on branches with the middle part of their feet, they should be mutually consistent with the arch of the branch surface, which might be the direct reasons of the formation of the longitudinal arches (Fig. 4B). The longitudinal arches are also helpful to keep the body on branches stably and rotate the body from the hypsokinesis posture to the anteverted posture, which is a necessary step in a two-arm brachiation for continuous forward locomotion. Our ancestors' Achilles tendon also evolved because of jumping or landing frequently on the branches.

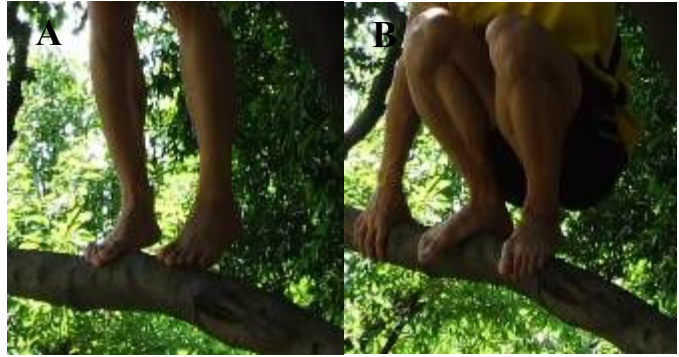

**Fig. 4.** Spatial positions of chimpanzee and human feet with branches. When ancient human jumped from, land to or stood and squatted on branches, their feet were usually perpendicular to the branches (**A**, **B**), accounting for longitudinal arches on their feet. Chimpanzees' feet of have no longitudinal arches because their feet are always parallel with branches.

We know that the longitudinal arch is important for running and jumping, but it initiated and formed by staying on and jumping from branches with feet perpendicular to tree branches. Jump and landing on the trees is difficult for modern human, but not for ancient human, because they had longer and stronger arms, longer toes, smaller bodies and they were expert in living in forest.

**Discussion**

Each kind of apes has its own mode of arboreal locomotion, and two-arm forward brachiation should be the most suitable one for our arboreal ancestors. During two-arm brachiation, when the center of gravity swings to the lowest position, the whole body was in a straightly erect posture, which is very similar to the standing posture of modern human.

It can be seen from the fossils of Australopithecus that their upper limbs were shorter than lower limbs, the proportion between the length of the upper limbs and that of the lower limbs is suitable for two-arm brachiation, but not suitable for any other kinds of quadruped walking.

There are two stages during the evolution of bipodalism, the first one is related to two-arm brachiation that can be dated back to about 6 million years ago, and the second one is related to hunting which may occurred about 2 million years ago. Hunting was important for the survival of ancient human, and one of the main reasons that ancient human chose hunting as one of the main means of obtaining food was that the main skills used in hunting, such as stick hitting and stone/spear throwing were the natural development from the two-arm forward brachiation. Although two million years ago, our ancestors might walk clumsily with wide but short steps, hereafter, larger heels and longer legs evolved as a result of long term walking and running, their walking posture evolved more and more like modern human.

Human evolution is complicated, there are many other factors coupled with two-arm brachiation facilitated ancient human to stand up and walk upright. But we think two-arm brachiation is the main cause of the evolution of human bipedalism, and our body structure especially the body joints are still affected by such arboreal locomotion mode, the corresponding knowledge could also help us to understand better human body and the inherited physiological factors that may be related to human's health, especially the physiological defects, weaknesses and diseases.

**Supplementary Materials:**

Movies S1-S3

Respectively, from website:

http://v.ku6.com/show/3eOJMLiL_wyrm-l2aoX1sg...html

http://v.ku6.com/show/gTMAqNDtfsQGK0yF0_Bjqg...html

http://video.sina.com.cn/v/b/113861422-3735632923.html